\documentclass[12pt]{article}
\usepackage{euler,epsfig,times,axodraw}
\begin{document}
\newcommand{\bq}{\begin{equation}}
\newcommand{\eq}{\end{equation}}
\newcommand{\bqa}{\begin{eqnarray}}
\newcommand{\eqa}{\end{eqnarray}}
\newcommand{\nl}{\nonumber \\}
\newcommand{\suml}{\sum\limits}
\newcommand{\bino}[2]{\left(\begin{tabular}{c} $#1$\\$#2$\end{tabular}\right)}
\newcommand{\f}{\varphi}
\newcommand{\plaat}[4]{\raisebox{#3cm}
      {\epsfig{figure=figs/#1.eps,width=#2cm,height=#4cm}}}
\newcommand{\diagr}[3]{\raisebox{#3cm}
      {\epsfig{figure=figs/#1.eps,width=#2cm}}}
\newcommand{\xver}{\;/\;}

\begin{center}
{\bf
{\Large Counting loop diagrams:\\
Computational complexity of\\ higher-order amplitude evaluation}\\
\vspace*{\baselineskip}
Ernst van Eijk\footnote{{\tt ernsteij@sci.kun.nl}}, 
Ronald Kleiss\footnote{{\tt kleiss@sci.kun.nl}}, 
Achilleas Lazopoulos\footnote{{\tt lazopoul@sci.kun.nl},\nl Research supported by the E.U. contract no. HPMD-CT-2001-00105}\\
HEFIN, University of Nijmegen, the Netherlands\\
\vspace*{3\baselineskip}
Abstract}
\end{center}
We discuss the computational complexity of the
perturbative evaluation of scattering amplitudes,
both by the Caravaglios-Moretti algorithm and by direct evaluation of
the individual diagrams. For a self-interacting scalar theory, we
determine the complexity as a function of the number of external legs.
We describe a method for obtaining the number of topologically
inequivalent Feynman graphs containing closed loops,
and apply this to one- and two-loop amplitudes. 
We also compute the number of
graphs weighted by their symmetry factors, thus arriving at exact and
asymptotic estimates for the average symmetry factor of diagrams. We present
results for the asymptotic number of diagrams up to 10 loops, and prove that
the average symmetry factor approaches unity as the number of external legs
becomes large.

\section{Introduction}
With the advent of high-energy colliders such as LHC and TESLA, 
high-multiplicity final states will become ever more relevant,
increasing the need for efficient evaluation of complicated
multi-leg amplitudes. Performing such calculations by a direct evaluation of
all relevant Feynman graphs is computationally hard in the sense that
the number of graphs increases with $N$ roughly as $N!$, the total 
number of external legs. For example, the $2\to8$ purely gluonic amplitude
in QCD contains 10.5 million Feynman graphs at the tree level; and
one may expect that loop corrections (described by many more
diagrams) will also be important. A computational breakthrough has been
achieved by the introduction of the Caravaglios-Moretti (CM)  algorithm \cite{CM},
in which the Schwinger-Dyson (SD) equations of the theory, rather
than their decomposition in individual Feynman diagrams, are employed,
thus leading to a complexity of order $c^N$ (where $c$ is a constant).
Such methods, however, have to date only be formulated for the Born 
approximation. Barring a revolutionary new method for solving the
SD equations including loop effects\footnote{In informal discussions, 
all the experts agree that this would be a tremendous advance --- but
no-one has a clue on how to approach it.}, the most straightforward
approach would seem to use the vertices of the effective, rather than those
of the bare, action. In such an approach the effective vertices with up to
$N$ legs have to be employed, which increases the complexity of the CM
algorithm. In order to assess the relative merit of the CM algorithm, it
is therefore relevant to compare the computational complexity of the
CM approach to the number of higher-order Feynman graphs.
In section 2, we calculate the number of individual diagrams,
not weighted by their symmetry factors, in zero-, one- and two-loop level for
four models of a self-interacting scalar theory. We also give the number of
one particle irreducible graphs, that is needed in the sequence. In section 3
we give the number of diagrams, now weighted by their symmetry factors, for the
four models, as they occur directly from the path integral. Section 4 contains asymptotic estimates,
in the number of external legs, for weighted and unweighted graphs. In section 5
we proceed in calculating the computational complexity of the CM algorithm in one and two
loops. In section 6 we compare the efficiency of the CM algorithm to that of the individual-diagram approach.

\section{Counting diagrams}
We consider a self-interacting scalar theory with arbitrary vertices
of the type $\f^k$, $k=3,4,\ldots$.
We define the `potential'
\begin{equation}
V(\f) = \suml_{k\ge3} \epsilon_k\;{\f^k\over k!}\;\;,
\end{equation}
where $\epsilon_k$ is 1 if the $\f^k$ interaction is present, otherwise it
is zero. We shall specialize to a number of cases:
\bqa
\mbox{$\f^3$ theory} &:& V(\f) = \f^3/6\;\;,\nl
\mbox{$\f^4$ theory} &:& V(\f) = \f^4/24\;\;,\nl
\mbox{gluonic QCD} &:& V(\f) = \f^3/6+\f^4/24\;\;,\nl
\mbox{effective theory} &:& V(\f) = e^{\f}-1-\f-\f^2/2\;\;;
\eqa
but alternative theories are easily implemented.

\subsection{Counting tree diagrams}
Tree diagrams can be conveniently counted by means of the SD equation. This
hinges on the fact that, at the tree level, all diagrams have unit symmetry
factor.
The counting of tree diagrams has been described in
detail in \cite{petroscount,petrosmultijet,petrosvertex},
and here we briefly recapitulate these results.
Let us denote by $a(n)$ the {\em number of Feynman tree diagrams\/}
contributing to the $1\to n$ amplitude, and define the generating
function
\begin{equation}
\phi_0(x) = \suml_{n\ge1}\;{x^n\over n!}\;a(n)\;\;.
\end{equation}
Pictorially, we denote this as
\begin{equation}
\phi_0(x) = \diagr{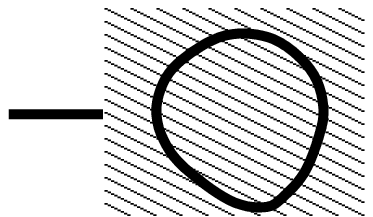}{2}{-0.5}
\end{equation}
By considering the alternatives when entering the blob from the left,
we easily see that
\begin{equation}
\phi_0(x) = \suml_{k\ge2}\;\epsilon_{k+1}\;\diagr{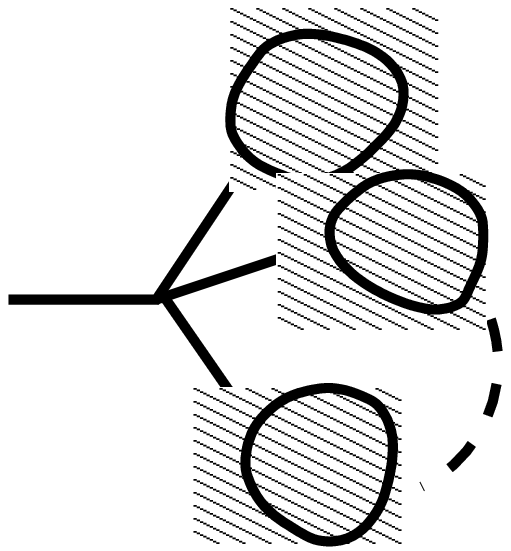}{2}{-1}\;\;,
\end{equation}
where the right-hand side contains $k$ blobs. This implies that $\phi_0(x)$
obeys the equation
\begin{equation}
\phi_0(x) = x + V'(\phi_0(x))\;\;.
\end{equation}
Since $V(\f)$ is of order ${\cal O}(\f^3)$, this SD equation can easily
be iterated starting with $\phi_0(x)=0$, and the desired $a(n)$ can be read off
once the iteration has proceeded far enough. Notice that
\begin{equation}
V''(\phi_0(x)) = 1 - {1\over\phi_0'(x)}\;\;\;,\;\;\;
V^{(p)}(\phi_0(x)) = {1\over\phi_0'(x)}
{d\over dx}V^{(p-1)}(\phi_0(x))\;\;(p\ge3)\;\;,
\end{equation}
so that the higher derivatives of $V(\phi_0(x))$ are completely expressed in
terms of $\phi_0'(x)$ and {\em its\/} derivatives.\\

The asymptotic behaviour of $a(n)$ for large $n$ is determined by the
singularity structure of $\phi_0(x)$. Since $\phi_0(x)$ cannot reach infinity
for finite values of $x$, the singularities take the form of branch cuts,
where $\phi_0(x)$ remains continuous but (as it turns out in all cases studied so
far) $\phi_0'(x)$ diverges. We have
\begin{equation}
x = \phi_0-V'(\phi_0)\;\;\;\Rightarrow\;\;\;
{dx\over d\phi_0} = 1 - V''(\phi_0)\;\;,
\end{equation}
and the dominant singularity is reached for that value $\phi_c$
for which $V''(\phi_c)=1$ and
\begin{equation}
x_c = \phi_c - V'(\phi_c)
\end{equation}
is closest to the origin\footnote{Here, we disregard the possibility that
there are several such values, arising from a symmetry of the potential such
as in the case of theories with only $\f^m$ interactions ($m\ge4$).
These cases are treated in detail in \cite{petrosthesis} and references
therein. The asymptotic results given here are `coarse-grained'.}. This value is
always located on the positive real axis,
where $\phi_0(x)$ is concave and monotically increasing for $x<x_c$.
Taylor expansion then gives the structure of the branch cut:
\begin{equation}
\phi_0(x) \sim \phi_c - \sqrt{{2x_c\over V^{(3)}(\phi_c)}}
\left(1-{x\over x_c}\right)^{1/2}\;\;,\label{approxim}
\end{equation}
from which we conclude that, for large $n$,
\begin{equation}
a(n) \sim \sqrt{{x_c\over2\pi V^{(3)}(\phi_c)}}\;\frac{n!}{ n^{3/2} x_c^{n}}=\frac{C}{\sqrt{4\pi}}\;{n!\over n^{3/2} x_c^{n-1/2}}\;\;.
\end{equation}
with $C\equiv \sqrt{2 / V^{(3)}(\phi_c)}$. In the table we give the relevant numbers for the four case theories.
\begin{center}\begin{tabular}{|l|c|c|c|}\hline\hline
theory & $\phi_c$ & $x_c$ & $C$ \\ \hline
$\f^3$    & 1             & $1/2$            & $\sqrt{2}$ \\
$\f^4$    & $\sqrt{2}$    & $\sqrt{8/9}$     & $2^{1/4}$ \\
gQCD      & $-1+\sqrt{3}$ & $\sqrt{3} - 4/3$ & $(4/3)^{1/4}$ \\
effective & $\log(2)$     & $2\log(2)-1$     & 1 \\
\hline\hline\end{tabular}\end{center}

\subsection{Counting one-loop diagrams}
When closed loops are introduced, an SD-type equation itself cannot be used to
count the number of {\em topologically inequivalent\/} graphs. This stems
from the fact that the SD-type equations are {\em local\/} in the sense
that they
only consider (in a recursive manner) what happens at a single vertex of
a diagram, while the topology of a graph containing closed loops is a
{\em global\/} property of the whole graph. Instead, one has to
settle for an order-by order and topology-by-topology treatment.\\

Every one-loop diagram can be viewed as a single closed loop, to which
tree-diagram pieces (which we call {\em leaves\/}) are attached. From
\begin{equation}
\diagr{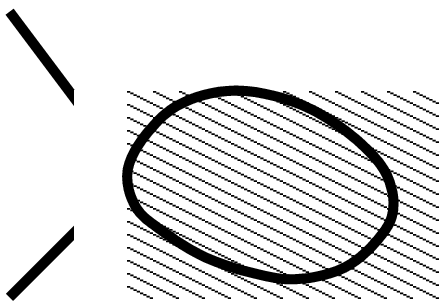}{1.5}{-.5} =
 \suml_{k\ge1}\epsilon_{k+2}\diagr{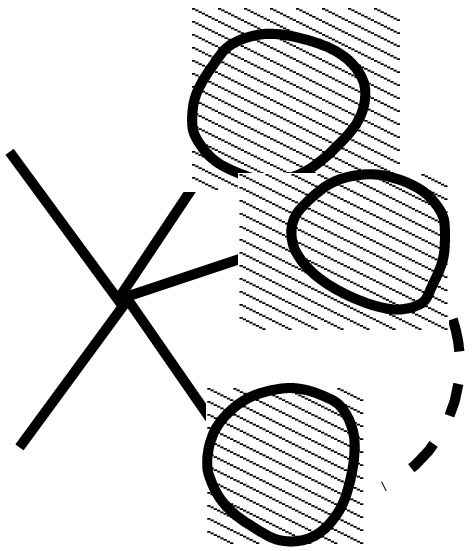}{2}{-1.} =
\suml_{k\ge1}{\epsilon_{k+2}\over k!}\phi_0(x)^k =
V''(\phi_0(x)) \equiv v\;\;,
\end{equation}
where the sum has $k$ blobs again, and we have
introduced the shorthand notation $v$, we see immediately that the
number of one-loop graphs can  be completely expressed in terms of $v$.
The generating function of $L_1(n)$, the
number of all one-loop non-vacuum graphs with precisely $n$ external legs,
is given by attaching leaves
to a closed loop in all possible ways:
\bqa
{\cal L}_1(x) &=& \suml_{n\ge1}{x^n\over n!}L_1(n) \nl
&=&
\diagr{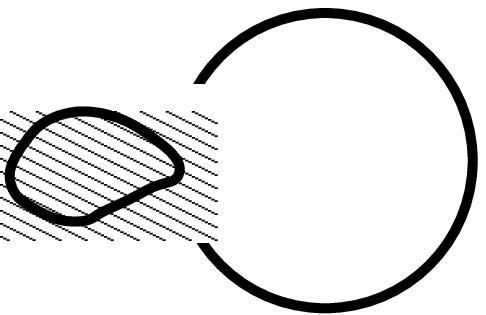}{1.2}{-.3} +
\diagr{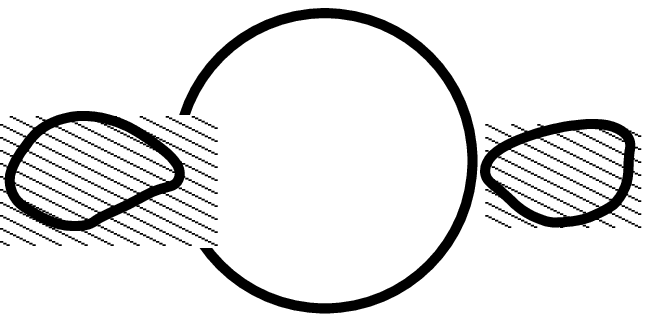}{1.5}{-.3} +
\diagr{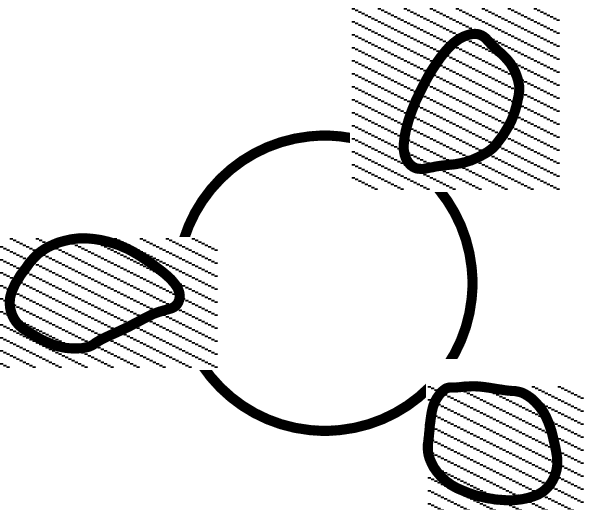}{1.5}{-.5} +
\diagr{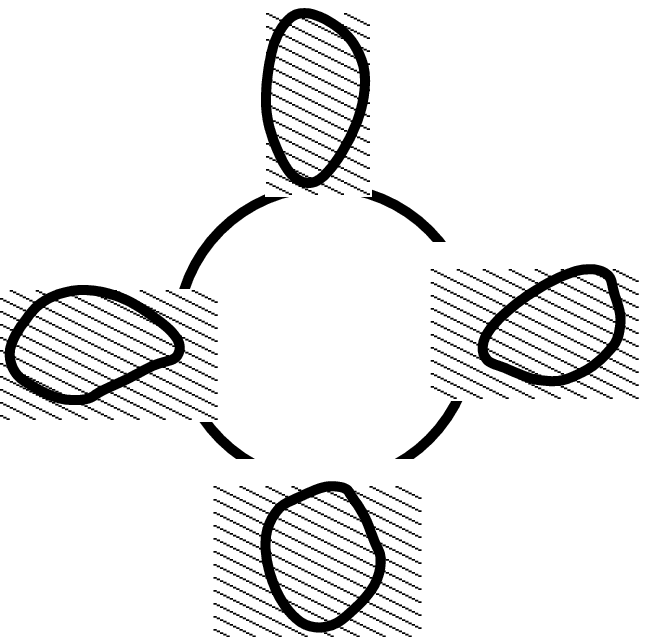}{1.5}{-.5} + \cdots
\label{eq:1lgraphs}
\eqa
The standard combinatorics for collecting the various external legs into
leaves, and inspecting the symmetry properties of the resulting graphs,
show that a one-loop graph with $m$ leaves has precisely the `natural'
symmetry factor $1/(2m)$, with two important exceptions: the graphs
with one or two leaves have an additional symmetry since, for the one-leave
graph, the loop line may be flipped over, and for the two-leave graph the
two internal loop lines may be interchanged. This leads us to the
strategy for computing the number of topologically inequivalent graphs:
\begin{itemize}
\item Write down the vacuum graphs, with their `natural symmetry factor';
\item Attach leaves in all possible places;
\item Multiply by the order of the residual symmetry left over after
 the particular attachment.
\end{itemize}
Performing this program for the one-loop case, we find
\bqa
{\cal L}_1(x) &=& {2\over2}v + {2\over4}v^2 + \suml_{m\ge3}{1\over2m}v^m
= {1\over2}v + {1\over4}v^2 - {1\over2}\log(1-v)\nl
&=&  {1\over2}v + {1\over4}v^2 + {1\over2}\log(\phi_0'(x))\;\;.
\eqa
The number of one-loop diagrams with $n$ external legs is given below
for some theories
\begin{center}
\begin{tabular}{|l|r|r|r|r|}\hline\hline
$N$ & $\f^3$ & $\f^4$ & gQCD & effective \\ \hline
1  & 1           & 0        & 1            & 1 \\
2  & 2           & 1        & 3            & 3 \\
3  & 7           & 0        & 14           & 15 \\
4  & 39          & 7        & 99           & 111 \\
5  & 297         & 0        & 947          & 1,104 \\
6  & 2,865        & 145      & 11,460        & 13,836 \\
7  & 33,435       & 0        & 167,660       & 209,340 \\
8  & 457,695      & 6475     & 2,876,580      & 3,711,672 \\
9  & 7,187,985     & 0        & 56,616,665     & 75,461,808 \\
10 & 127,356,705   & 503,440   &  1,257,154,920  & 1,730,420,592 \\
\hline\hline
\end{tabular}
\end{center}

\subsection{Counting two-loop diagrams}
At the two-loop level, there are three topologically different vacuum
diagrams. These are:
\begin{equation}
\mbox{{\bf a:}}\;\diagr{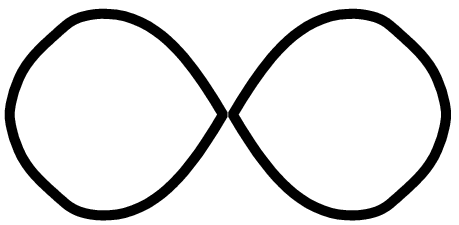}{2}{-.5}\;{1\over8}\;\;\;,\;\;\;
\mbox{{\bf b:}}\;\diagr{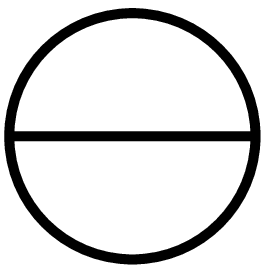}{1.3}{-.5}\;{1\over12}\;\;\;,\;\;\;
\mbox{{\bf c:}}\;\diagr{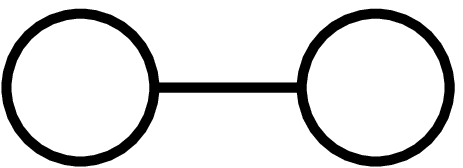}{2}{-.3}\;{1\over8}\;\;\;,\;\;\;
\label{eq:2lgraphs}
\end{equation}
where we have indicated their `natural' symmetry factor.
Since these graphs contain vertices, we must also accommodate leaves
attaching themselves to vertices:
\begin{equation}
\diagr{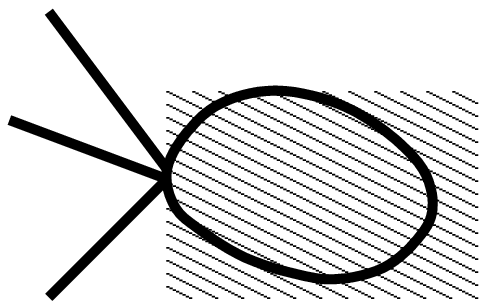}{1.5}{-.5} = V^{(3)}(\phi(x))-V^{(3)}(0)\;\;\;,\;\;\;
\diagr{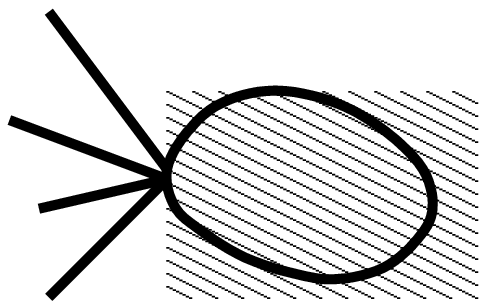}{1.5}{-.5} = V^{(4)}(\phi(x))-V^{(4)}(0)\;\;.
\end{equation}
In case no leave happens to be attached, the expression for the vertices
read, of course, $V^{(3)}(0)$ and $V^{(4)}(0)$, respectively. This
prohibits, for instance, the occurrence of a three-point vertex
in a $\f^4$ theory.
For each of the graphs we have to admit the possibility of zero, one, or more
leaves on each line,
and that of leaves on any vertex. For the determination of
the residual symmetries it must be remembered that lines without leaves
on them {\em may\/} be interchanged, and vertices without leaves {\em may\/}
be interchanged, provided the `anchoring' of the graph to the external
legs contained in every leaf present permits such an interchange. As
a simple example, the vacuum graphs themselves, without any leaves on them,
have a residual symmetry of precisely 8,12, and 8, respectively, so that
indeed they will be counted precisely one time. For graph {\bf a}
there are now $2\times3^2=18$ cases to be considered, and for {\bf b} and
{\bf c} we have $2^2\times3^3=108$ cases. The results for their
generating functions are:
\bqa
{\cal L}_2^{\mbox{{\bf(a)}}}(x) &=&
{1\over8}V^{(4)}(\phi)\left[
\left(1+v+{1\over1-v}\right)^2+4\right]\;\;,\nl
{\cal L}_2^{\mbox{{\bf(b)}}}(x) &=&
{1\over12}\left[
\left(V^{(3)}(\phi)\right)^2\left(2+{3\over1-v}+{1\over(1-v)^3}\right)
\right.\nl & & \left.\hspace{1cm}
+\left(V^{(3)}(0)\right)^2\left(2+3(1+v)+(1+v)^3\right)\right]\;\;,\nl
{\cal L}_2^{\mbox{{\bf(c)}}}(x) &=&
{1\over8}\left[\left(V^{(3)}(\phi)\right)^2
{1\over1-v}\left(1+v+{1\over1-v}\right)^2
\right.\nl & & \left.\hspace{1cm}
+4\left(V^{(3)}(0)\right)^2(1+v)\right]\;\;.
\eqa
The total number $L_2(n)$ of two-loop graphs with precisely $n$
external lines is therefore given via
\begin{equation}
{\cal L}_2(x) = \suml_{n\ge0}{x^n\over n!}L_2(n) =
{\cal L}_2^{\mbox{{\bf(a)}}}(x) +
{\cal L}_2^{\mbox{{\bf(b)}}}(x) +
{\cal L}_2^{\mbox{{\bf(c)}}}(x)\;\;.
\end{equation}
Below, we give again the results for our specific theories.
\begin{center}
\begin{tabular}{|l|r|r|r|r|}\hline\hline
$N$ & $\f^3$ & $\f^4$ & gQCD & effective \\ \hline
0  & 2             & 1          & 3              & 3\\
1  & 3             & 0          & 6              & 7\\
2  & 10            & 3          & 29             & 35\\
3  & 58            & 0          & 217            & 273\\
4  & 465           & 42         & 2,214           & 2,876\\
5  & 4,725          & 0          & 28,365          & 38,034\\
6  & 57,900         & 1,485       & 436,780         & 604,320\\
7  & 829,080        & 0          & 7,847,420        & 11,202,156\\
8  & 13,570,515      & 97,335      & 161,048,720      & 237,187,552\\
9  & 249,789,015     & 0          & 3,715,400,500     & 5,645,523,408\\
10 & 5,105,239,650    & 10,210,200   & 95,156,789,700    & 149,180,360,320\\
\hline\hline
\end{tabular}\end{center}
\vspace*{10mm}

The extension to three or more loops is a matter of establishing
the vacuum diagrams. For the three-loop case,
however, there are 15 such graphs. Dressing them with leaves leads to
a larger number of cases to be considered,
ranging from 54 to 11,664 per graph.

\subsection{Counting amputated diagrams}
Loop diagrams containing tadpoles or seagulls are constant contributions to lower order diagrams
and are usually ignored. Moreover, diagrams containing self-energy loops on external legs are absorbed, during the
renormalization procedure. Removing such diagrams from the above results is a simple task. One has to substract all contributions
from (a) diagrams with loops carrying zero or one vertex and (b) diagrams carrying two vertices one of which is connected with an
external leg while the other is a single propagator.

For the one-loop case one has to substract the first graph in eq.~\ref{eq:1lgraphs}, as a set of tadpole or seagul diagrams,
as well as a contribution from graphs of the form
\begin{center}
\begin{picture}(100,30)(0,0)
\Line(0,15)(20,15)
\BCirc(35,15){15}
\Line(50,15)(70,15)
\GCirc(80,15){10}{0.7}
\end{picture}
\end{center}

With these modifications the generating function reads

\begin{equation}
{\cal L}_1(x) = -{1\over2}v + {1\over4}v^2 - {1\over2}\log(1-v)-x \phi_0(x) V^{(3)}(0)
\end{equation}

The number of amputated one-loop diagrams for our test theories is given below:

\begin{center}
\begin{tabular}{|l|r|r|r|r|}\hline\hline
$N$ & $\f^3$ & $\f^4$ & gQCD & effective \\ \hline
3  & 1           & 0        & 4           & 4 \\
4  & 12          & 3        & 39           & 43 \\
5  & 117         & 0        & 437          & 502 \\
6  & 1,290        & 75      & 5,800        & 6,916 \\
7  &  16,425      & 0        & 90,450       & 111,660 \\
8  & 239,400      & 3675     & 1,627,640      & 2,077,944 \\
9  & 3,944,745     & 0        & 33,258,715     & 43,883,696 \\
10 & 72,627,030   & 303240   &  761,405,820  & 1,037,955,824 \\
\hline\hline
\end{tabular}
\end{center}

For two-loops diagrams one has to consider separately each vaccuum graph.
All graphs containing loops with less than two vertices should be removed, as well as a variety of special cases
which lead to non-amputated diagrams.

The generating functions for each of the three basic topologies becomes

\bqa
{\cal L}_2^{\mbox{{\bf(a)}}}(x) &=&{1\over8}V^{(4)}(\phi)\left(\frac{v^2(2-v)^2}{(1-v)^2}\right)-V^{(4)}(0)V^{(3)}(0)x\phi_0(x)\;\;\nl
{\cal L}_2^{\mbox{{\bf(b)}}}(x) &=&{1\over12} \left[ \left(V^{(3)}(\phi_0)\right)^2 \left(\frac{6-12v+9v^2-2v^3}{(1-v)^3}\right) \right]\;\;\nl
& &-\frac{1}{12}\left(V^{(3)}(0)\right)^2 \left[ 12x\phi_0+\frac{6-3v^2+2v^3+v^4}{1-v} \right]\;\;\nl
& &-\left(V^{(4)}(0)\right)^2x\phi_0(x)-V^{(3)}(0)V^{(3)}(\phi_0)-V^{(3)}(0)V^{(4)}(0) 2 x\phi_0(x)\;\;,\nl
{\cal L}_2^{\mbox{{\bf(c)}}}(x) &=&\frac{1}{8}\left[\left(V^{(3)}(\phi)\right)^2   \frac{v^2(2-v)^2}{(1-v)^3} \right]-\frac{1}{8}V^{(3)}(\phi)V^{(3)}(0) x \frac{4v(2-v)}{(1-v)^2}\;\;\nl
& &+\left(V^{(3)}(0)\right)^2 \frac{4}{8} \frac{x^2}{1-v}
\eqa

The exact number of two-loop connected amputated diagrams for our test theories is given below

\begin{center}
\begin{tabular}{|l|r|r|r|r|}\hline\hline
$N$ & $\f^3$ & $\f^4$ & gQCD & effective \\ \hline
3  & 4           	& 0        & 28           & 37 \\
4  & 63          	& 9        & 457           & 600 \\
5  & 870         	& 0        & 7,285          & 9,760 \\
6  & 12,945      	& 460      & 128,675        & 177,160 \\
7  &  212,940    	& 0        & 2,552,165       & 3,617,824 \\
8  & 3,874,815   	& 35,315     & 56,538,055      & 82,588,784 \\
9  & 77,605,290  	& 0        & 1,387,411,690     & 2,089,438,256 \\
10 & 1,700,078,625	& 4,090,800   &  37,407,699,175  & 58,096,995,744 \\
\hline\hline
\end{tabular}
\end{center}

\subsection{Counting 1PI diagrams}
The same methods as above can easily be employed in order to
compute the number of one-particle irreducible (1PI) diagrams. We simply
restrict ourselves to the 1PI vacuum bubbles; and, since 1PI diagrams
cannot have any vertex in their leaves, we simply replace
$\phi_0(x)$ in the arguments of $V'',V^{(3)},V^{(4)},\ldots$ by $x$.
For the generating function of the 1PI one-loop diagrams, we therefore
have
\begin{equation}
{\cal L}^{\mbox{\small 1PI}}_1(x) =
{1\over2}w + {1\over4}w^2 - {1\over2}\log(1-w)\;\;\;,\;\;\;
w = V''(x)\;\;.
\eq
The resulting numbers are given in the following table.
\begin{center}
\begin{tabular}{|l|r|r|r|r|}\hline\hline
$N$ & $\f^3$ & $\f^4$ & gQCD & effective \\ \hline
1  & 1        & 0      & 1         & 1 \\
2  & 1        & 1      & 2         & 2 \\
3  & 1        & 0      & 4         & 5 \\
4  & 3        & 3      & 12        & 17 \\
5  & 12       & 0      & 57        & 83 \\
6  & 60       & 15     & 390       & 557 \\
7  & 360      & 0      & 3,195      & 4,715 \\
8  & 2,520     & 315    & 30,555     & 47,357 \\
9  & 20,160    & 0      & 333,900    & 545,963 \\
10 & 181,440   & 11,340  & 4,105,080   & 7,087,517 \\
\hline\hline\end{tabular}\end{center}

At the two-loop level, we similarly find
\bqa
{\cal L}^{\mbox{\small 1PI}}_2(x) &=&
{1\over8}V^{(4)}(x)\left[
\left(1+w+{1\over1-w}\right)^2+4\right]\nl
& & + {1\over12}V^{(3)}(x)^2\left[2+{3\over1-w}+{1\over(1-w)^3}\right]\nl
& & + {1\over12}V^{(3)}(0)^2\left[2+3(1+w)+(1+w)^3\right]\;\;.
\eqa
Numbers are given below.
\begin{center}
\begin{tabular}{|l|r|r|r|r|}\hline\hline
$N$ & $\f^3$ & $\f^4$ & gQCD & effective \\ \hline
0  & 1          & 1        & 2            & 2 \\
1  & 1          & 0        & 3            & 4 \\
2  & 2          & 2        & 9            & 13 \\
3  & 7          & 0        & 40           & 62 \\
4  & 36         & 12       & 265          & 410 \\
5  & 240        & 0        & 2,230         & 3,499 \\
6  & 1,860       & 225      & 22,485        & 36,213 \\
7  & 16,380      & 0        & 261,135       & 435,852 \\
8  & 161,280     & 8,295     & 3,418,695      & 5,944,000 \\
9  & 1,753,920    & 0        & 49,712,670     & 90,309,029 \\
10 & 20,865,600   & 481,950   & 794,102,400    & 1,510,208,963 \\
\hline\hline\end{tabular}\end{center}

\section{Counting with symmetry factors}
The counting of diagrams including their symmetry factors is a somewhat
simpler task, which can be performed on the basis of the path integral  itself. 
In ~\cite{cvitanovic} this has been discussed in detail. However our approach here is somewhat different.
One can expand the generating function of the number of connected diagrams
perturbatively around $\varphi=0$ and get a series in $x$ (the source). Or, alternatively,
one can expand perturbatively around the tree level one-point function $\varphi=\phi_0$.
This shift eliminates the source $x$ in favour of the tree level one-point function
$\phi_0(x)$, and reveals the vaccuum graph dressing procedure
that we employed above.
\subsection{Counting diagrams with symmetry factors}
Consider the generating function for the number of disconnected diagrams of a
scalar theory with arbitrary couplings and a source $x$:
\begin{equation}
Z(x)=N\int d\varphi \;\;exp\left(-\frac{1}{\hbar}(\frac{1}{2}\varphi^2-V(\varphi)+x\varphi)\right)
\end{equation}
with $N=1/\sqrt{2 \pi \hbar}$. Expanding around the tree level approximation $\phi_0$ of the one-point function, i.e. setting $\varphi\rightarrow
\phi_0+\varphi$, and making use of the Schwinger -Dyson equation for $\phi_0(x)$ gives
\begin{equation}
Z(x)=N\;\;exp(-\frac{1}{\hbar}S(\phi_0)+\frac{x\phi_0}{\hbar}) \int d \varphi \;e^{-\frac{1}{\hbar}\hat{S}(\varphi)}
\end{equation}
with
\begin{equation}
\hat{S}(\varphi)=\frac{1-V^{\prime\prime}(\phi_0)}{2}\varphi^2-\sum_{n=3}^{\infty}V^{(n)}(\phi_0)\frac{\varphi^n}{n!}
\end{equation}
The generating function of the number of connected diagrams is then
\begin{equation}
\label{eq:W(x)}
W(x)=\hbar \log(Z(x))=-S(\phi_0)+x\phi_0 +\hbar \log(N\int d \varphi e^{-\frac{1}{\hbar}\hat{S}(\varphi)})
\end{equation}
 We see that it can be seen as a sum of the tree
level part plus higher order corrections. These corrections can be written as the generating function
for the vacuum diagrams of a theory with
action  $\hat{S}(\varphi)$. The Feynman rules corresponding to this
action can be read off directly :
\begin{itemize}
\item $\frac{1}{1-V^{\prime\prime}(\phi_0)}=\phi^{\prime}_0$ for every propagator.
\item $V^{(n)}(\phi_0)$ for every n-point vertex.
\end{itemize}

Given the potential $V(\varphi)$ of the theory one can expand the vertex terms
 in the exponential of eq.~\ref{eq:W(x)},
calculate the Gaussian integrals and arrive at an expression for $W(x)$ that
 contains only $V^{\prime\prime}(\phi_0(x))$ and its derivatives.
 In this
way, given the tree level one-point function of the theory, one finds the number
 of graphs weighted by their symmetry factors
to arbitrary order.

Writing $W(x)$ in an $\hbar$ expansion
\begin{equation}
W(x)=W_0(x)+\hbar W_1(x)+\hbar^2 W_2(x)+\ldots
\end{equation}
and, performing the integral and collecting together the terms of the same order in $\hbar$, we see that the one loop
diagrams are generated by
\begin{equation}
W_1(x)=\frac{1}{2}\log(\frac{1}{1-V^{\prime\prime}(\phi_0)})
\end{equation}

We can also find
the generating function for the two loop diagrams
\begin{equation}
W_2(x)=\frac{1}{8}\frac{V^{(4)}(\phi_0)}{(1-V^{\prime\prime}(\phi_0))^2}+\frac{5}{24}\frac{V^{(3)}(\phi_0)V^{(3)}(\phi_0)}{(1-V^{\prime\prime}(\phi_0))^3}
\end{equation}
The factor $\frac{1}{8}$ in front of the first term is the symmetry factor of the
only 2-loop vacuum diagram with a 4-vertex (see fig ~\ref{eq:2lgraphs}.a). The factor $\frac{5}{24}=\frac{1}{8}+\frac{1}{12}$
is the sum of the symmetry factors of the two vacuum diagrams with two 3-vertices
(see fig ~\ref{eq:2lgraphs} .b and .c)\footnote{In fact  one could even avoid performing the integral since the generating function for N loops
is simply the sum of the vacuum graphs with
 N loops weighted by their symmetry factors using the Feynman rules for
 the $\hat{S}(\varphi)$ action given above. However, this presupposes that one knows
 what the symmetry factor of the specific vacuum diagram is.}.

Writing the derivatives  $V^{(m)}(\phi_0)$ in terms of derivatives of $\phi_0$ (which can be done by differentiating
the Schwinger-Dyson equation for $\phi_0$) one arrives at
\begin{equation}
W_2(x)=\frac{1}{8}\frac{\phi_0^{\prime\prime\prime}}{(\phi_0^{\prime})^2}-\frac{1}{6}\frac{(\phi_0^{\prime\prime})^2}{(\phi_0^{\prime})^3}
\end{equation}

Below, we give results for our four case theories in 1 loop.
\begin{center}
\begin{tabular}{|l|r|r|r|r|}\hline\hline
$N$ & $\f^3$ & $\f^4$ & gQCD & effective \\ \hline
1  & 1/2         & 0        & 1/2           & 1/2\\
2  & 1           & 1/2      & 3/2           & 3/2\\
3  & 4           & 0        & 15/2          & 8\\
4  & 24          & 7/2      & 57            & 63\\
5  & 192         & 0        & 1,149/2        & 658\\
6  & 1,920        & 80       & 7,230          & 8,568\\
7  & 23,040       & 0        & 218,175/2      & 133,676\\
8  & 322,560      & 3,815     & 1,919,190       & 2,430,816\\
9  & 5,160,960     & 0        & 77,146,125/2    & 50,484,016\\
10 & 92,897,280    & 31,0940   & 871,927,770     & 1,178,963,856\\
\hline\hline
\end{tabular}\end{center}

The results for the four theories in 2 loops are again collected below.
\begin{center}
\begin{tabular}{|l|r|r|r|r|}\hline\hline
$N$ & $\f^3$ & $\f^4$ & gQCD & effective \\ \hline
1  & 5/8              & 0            & 31/24             & 17/12\\
2  & 25/8             & 2/3          & 25/3              & 19/2\\
3  & 175/8            & 0            & 1,777/24           & 527/6\\
4  & 1,575/8           & 149/12       & 5,057/6            & 1,037\\
5  & 17,325/8          & 0            & 280,735/24         & 44,726/3\\
6  & 225,225/8         & 1,535/3       & 1,149,515/6         & 252,734\\
7  & 3,378,375/8        & 0            & 86,813,545/24       & 14,808,232/3\\
8  & 57,432,375/8       & 111,755/3     & 464,096,885/6       & 109,143,424\\
9  & 1,091,215,125/8     & 0            & 44,344,732,495/24
& 8,085,390,392/3\\
10 & 22,915,517,625/8    & 12,672,800/3   & 292,590,237,275/6
& 73,514,104,288\\
\hline\hline\end{tabular}\end{center}
Both in the one and two loop cases an intriguing pattern of denominators
is apparent for large $N$ values, which seems to persist (we have checked this
for $N$ up to 50).\\

The above procedure can easily be extended to higher-loop
amplitudes as well, but since we have not computed the unweighted diagram sums
we defer this discussion to the case of asymptotically large $N$.

\subsection{counting 1PI graphs}
The generating function for the one particle irreducible diagrams of a theory weighted
by their symmetry factors can be obtained by the same prescription
by substituting $\phi_0=x$. Now, however, we have to take into account only the
1PI vacuum diagrams. In the one loop case the only vacuum graph is 1PI
and the generating function is
\begin{equation}
W_1(x)=\frac{1}{2}\log(\frac{1}{1-V^{\prime\prime}(x)})
\end{equation}
In the two loop case we have to take into account the vacuum graph with one 4-vertex
(see figure ~\ref{eq:2lgraphs}.a) and only one of the two vacuum graphs
with three vertices (see figure ~\ref{eq:2lgraphs}.b) since the
other one (see figure ~\ref{eq:2lgraphs}.c) is not 1PI . This alters the symmetry factor from $\frac{5}{24}$
to $\frac{1}{12}$. We get then
\begin{equation}
W_2(x)=\frac{1}{8}\frac{V^{(4)}(x)}{(1-V^{\prime\prime}(x))^2}+\frac{1}{12}\frac{V^{(3)}(x)V^{(3)}(x)}{(1-V^{\prime\prime}(x))^3}
\end{equation}
We give below the number of irreducible diagrams weighted by their symmetry factors
in the 1-loop case for the four test theories :

\begin{center}
\begin{tabular}{|l|l|l|l|l|l|}
\hline \hline
  & $\varphi^3$ & $\varphi^4$ & $\varphi^3+\varphi^4$ & effective  \\
\hline
N=1 	& 1/2			& 0			& 1/2			& 1/2		\\
N=2  	& 1/2 			& 1/2 		& 1 				& 1 			 \\
N=3  	& 1 				& 0		 	& 5/2 			& 3 			 \\
N=4  	& 3 				& 3/2 		& 21/2 			& 13 		\\
N=5  	& 12 			& 0 		 	& 57 			& 75 		\\
N=6  	&  60			& 15			& 390			& 541 		\\
N=7         & 360			& 0			&3,195			&4,683		\\
N=8		&2,520			&315		&30,555			&47,293		\\
N=9		&20,160			&0			&333,900			&545,853		\\
N=10	&181,440			&11,440		&4,105,080		&708,7261	\\
\hline
\hline
\end{tabular}
\end{center}

We give below the number of irreducible diagrams weighted by their symmetry factors
in the 2-loop case for the four test theories :

\begin{center}
\begin{tabular}{|l|l|l|l|l|l|}
\hline \hline
  & $\varphi^3$ & $\varphi^4$ & $\varphi^3+\varphi^4$ & effective  \\
\hline
N=1 	& 1/4			& 0			& 2/3			& 19/24		\\
N=2  	& 1 				& 5/12 		& 41/12 				& 101/24 			 \\
N=3  	& 5 				& 0		 	& 89/4 			& 691/24 			 \\
N=4  	& 30 			& 21/4 		& 709/4 			& 5765/24 		\\
N=5  	& 210 			& 0 		 	& 1,660 			& 56,659/24 		\\
N=6  	&  1,680			& 135		& 17,865			& 64,0421/24 		\\
N=7         & 15,120			& 0			&217,035			&8,178,931/24		\\
N=8		&151,200			&5,775		&2,936,745			&116,422,085/24		\\
N=9		&1,663,200		&0			&43,787,520			&1,827,127,699/24		\\
N=10	&19,958,400		&368,550		&713,163,150		&31,336,832,741/24	\\
\hline
\hline
\end{tabular}
\end{center}

\section{Asymptotic estimates}
It is fairly easy to estimate the number of diagrams, both with and without
their symmetry factors, for asymptotically large $N$.
As before, the asymptotic behaviour of these numbers is governed by the
analytic structure of their generating functions close to that singularity
which is closest to the origin (that is, around $x\sim x_c$).
There, we have
\begin{equation}
\phi_0'(x) \sim {1\over2}C\left(x_c-x\right)^{-1/2}\;\;\;,\;\;\;
C = \left(2/V^{(3)}(\phi_c)\right)^{1/2}\;\;,
\end{equation}
where $x_c$, $\phi_c$ and $C$ again depend on the theory. Let us first
concentrate on the one-loop diagrams. Since $v=1-1/\phi_0'(x)$ has a
square-root branch cut at the singular point, $\log(1-v)$ is more
singular than $v$ or $v^2$, and we have
\begin{equation}
{\cal L}_1(x) \sim {1\over2}\log\left({C\over2\sqrt{x_c-x}}\right)
= {\cal L}_1^{(s)}(x)\;\;.
\end{equation}
We conclude that, for one-loop diagrams, the {\em average symmetry factor}
of a given diagram is asymptotically equal to 1.
The number $K_1(N)$ of graphs contained in the one-loop $N$-point
amplitude is asymptotically given by
\begin{equation}
K_1(N) \sim {1\over4}{1\over(x_c)^N}{N!\over N}
\end{equation}
To illustrate the convergence of the weighted number of graphs to the
unweighted number, we give the ratio of the coefficients of $x^N$ in
${\cal L}_1^{(s)}(x)$ to those of ${\cal L}_1(x)$ as a function of $N$ below,
for the pure $\f^3$ theory. The other cases show a similar
behaviour\footnote{For the pure $\f^4$ theory, this holds in the
`coarse-grained' approximation \cite{petrosthesis}.}, in which the asymptotic
regime is approached as $1/\sqrt{N}$: this can also be easily checked from
the exact form of ${\cal L}_1(x)$ close to the singularity.
\begin{center}
$$\plaat{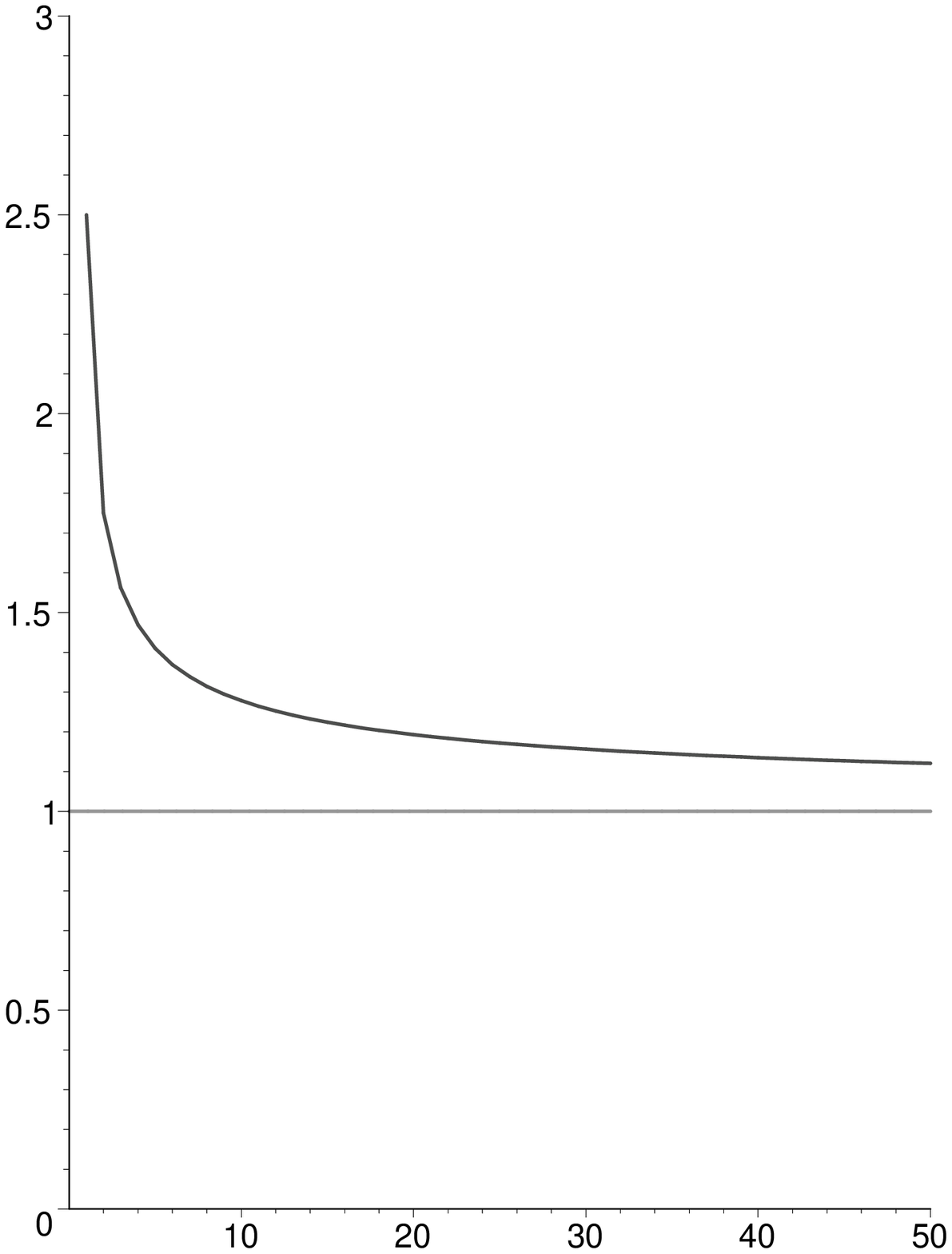}{8}{0}{8}$$
\end{center}
The asymptotic results for the higher-loop amplitudes can be established
by the following reasoning. The leading contribution from each leave-dressed
vacuum diagram is given by that part that has the highest degree of divergence
as $x\to x_c$. From each line in the vacuum graph, this is a factor
$1/(1-v) = \phi_0'(x)$. Furthermore, from each
$k$-point vertex in the vacuum graph the leading contribution
comes from the limiting behaviour of $V^{(k)}(\phi_0(x))$. Now, it is
easily seen that, as $x\to x_c$,
\begin{equation}
V^{(3)}(\phi_0(x)) \sim {2\over C^2}\;\;\;\Rightarrow
V^{(k)}(\phi_0(x)) \sim 0\;\;,\;\;k\ge4\;\;.
\end{equation}
We conclude that the leading behaviour of the number of unweighted
graphs is given by those vacuum graphs that contain only three-point
vertices. To get the number of unweighted diagrams at the $L$-loop level,
therefore, we first compute the normalized path integral for the pure
$\f^3$ theory, using the usual perturbative interchange between
expansion of the potential term and integration:
\bqa
Z &=& N\int\limits_{-\infty}^\infty d\f\;\exp\left(
-{\mu\over2}\f^2 + {\lambda\over6}\f^3\right)\nl
&=& \suml_{n\ge0}\;{(6n)!\over(2n)!(3n)!(576)^n}
\left({\lambda^2\over\mu^3}\right)^n\;\;.
\eqa
The sum of all connected vacuum diagrams with interactions is
then given by
\begin{equation}
W = \log(Z)\;\;,
\end{equation}
in the expansion of which the $L$-loop contribution ($L\ge2$) is given by
the term with $\lambda^{2L-2}$.
In this expression, it suffices to replace $\lambda$ by $2/C^2$ and
$\mu$ by $1/\phi_0'(x)$. The result is
\begin{equation}
W = \suml_{L\ge2}\;w_L\;C^{1-L}\;(x_c-x)^{3(1-L)/2}\;\;.
\end{equation}
The first coefficients $w_L$ are given below.
\begin{center}\begin{tabular}{|r|r|r|r|}\hline\hline
$L$ & $w_L$ & $L$ & $w_L$ \\ \hline
2 & ${5\xver48}$        & 7  & ${19675\xver6144}$              \\
3 & ${5\xver64}$        & 8  & ${1282031525\xver88080384}$      \\
4 & ${1105\xver9216}$   & 9  & ${80727925\xver1048576}$         \\
5 & ${565\xver2048}$    & 10 & ${1683480621875\xver3623878656}$ \\
6 & ${82825\xver98304}$ & 11 & ${13209845125\xver4194304}$      \\
\hline\hline\end{tabular}\end{center}
The asymptotic result for $K_L(N)$,
the number of unweighted diagrams contributing
to the $L$-loop $n$-point amplitude is therefore given by
\begin{equation}
K_L(N) \sim {\Gamma\left(N+{3\over2}(L-1)\right)\over
\left(x_c^{3/2}C\right)^{L-1}\Gamma\left({3\over2}(L-1)x_c^N\right)}\;\;.
\end{equation}
For the number of $L$-loop graphs weighted by their symmetry factors we
may employ the following formulation of the SD equation:
\begin{equation}
\phi_{L} = \suml_{\{n_{p,q}\}\ge0}V^{(m)}
\prod\limits_{p,q\ge0}\;{1\over(n_{p,q})!}
\left({1\over(q+1)!}\phi_p^{(q)}\right)^{n_{p,q}}\;\;,\label{SDform}
\end{equation}
where the bracketed superscripts denote derivatives, and
\begin{equation}
\suml_{p,q}\;(p+q)n_{p,q} = L\;\;\;,\;\;\;m = 1+\suml_{p,q}(q+1)n_{p,q}\;\;.
\end{equation}
The successive expressions for $\phi_L(x)$ in terms of lower-loop ones
can straightforwardly be worked out. For $L=1,2$ these have been given
in the previous section. If we now put in the approximate form of $\phi_0(x)$
given in Eq.(\ref{approxim}), it is easily checked (at least up to $L=10$)
that expression $W$ is reproduced. Note that in this approximation the
fourth and higher derivatives of $V(\phi_0)$ vanish, so that
Eq.(\ref{SDform}) is actually more complicated than need be: nevertheless,
by using the next-to-leading expression
\begin{equation}
\phi_0(x) \sim \phi_c - C(x_c-x)^{1/2} - C'(x_c-x)\;\;,
\end{equation}
it can also be checked that, indeed, the subleading  behaviour of
$\phi_0(x)$ shows up only in the subleading terms in $K_L(N)$. We conclude
that {\em as $N\to\infty$, the average symmetry factor of any Feynman
diagram approaches unity}.

\section{Complexity of the Caravaglios-Moretti algorithm}
\subsection{introduction}
The CM algorithm, as first explicitly given in \cite{CM} (and earlier
implied by \cite{bg}), consists of the computation of subamplitudes with
one off-shell leg, the other legs corresponding to on-shell
external legs of the transition matrix element. For tree diagrams, these
subamplitudes can be unambiguously specified by the particular set of external
momenta involved because of momentum conservation. For detailed descriptions,
we refer to \cite{CM,petrosthesis,multigluon}: here, we are only
interested in the combinatorics of the algorithm.

\subsection{Complexity for tree level computations in any theory}
We assume an $N$-particle
process, and set $K=N-1$. Each subamplitude can then be encoded by a binary
string with $N$ bits, each referring to a given external particle. The bit
is set to 1 if its external leg is involved in the subamplitude, and to
0 otherwise. For instance, the string $(1,1,0,1,1,0,0,0,\ldots,0,0)$ denotes
that subamplitude in which the external particles with labels 1,2,4 and 5
are combined, using the vertices of the theory, into a single off-shell
momentum.
By the same convention, a string with a single 1 refers to the Feynman rule
for a single external particle (a spinor or antispinor for fermions,
a polarization vector for vector particles, etctera).
The CM algorithm combines subamplitudes into successively more
complicated ones, culminating in the string $(1,1,1,\ldots,1,1,1,0)$,
which, after multiplying with the external factor $(0,0,0,\ldots,0,0,0,1)$
gives the final answer for the amplitude. It is clear that of the $N$
external particles, one can be left out of the combinatorics since it has to
be included only at the very end. The combinatorial problem is, therefore,
to determine the number of ways to decompose a string of $K$ bits. An
example of a possible decomposition is
$$
\plaat{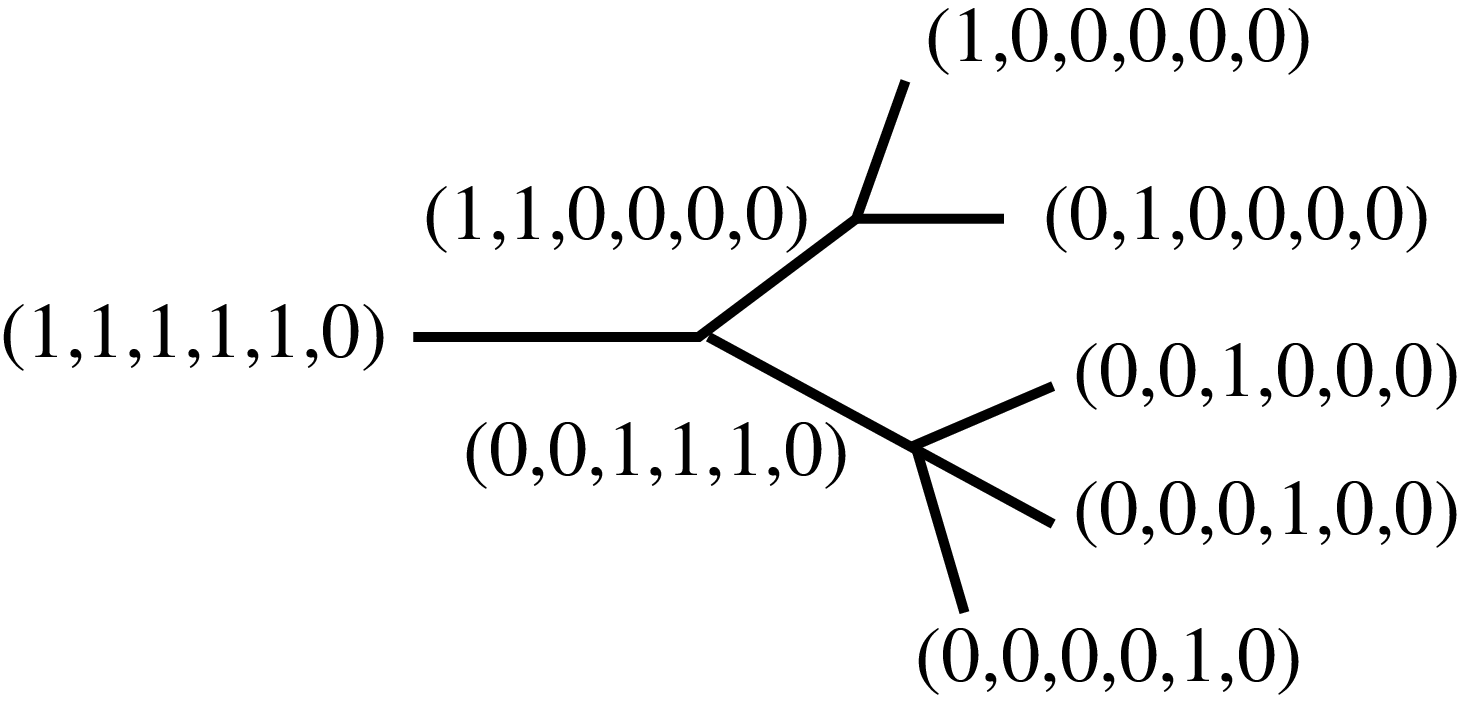}{10}{0}{4}
$$
where we have indicated the strings corresponding with the external legs
and the various subamplitudes. The possible decompositions depend on the
theory in question: the presence of an $(m+1)$-point vertex in the theory
allows for a decomposition into $m$ smaller strings. In this paper, we
shall only deal with theories of a single self-interacting field (gluonic
QCD being an example): extensions to more fields are fairly straightforward.
In recent implementations such as \emph{HELAC} (\cite{HELAC}), this decomposition can be recognised explicitly.\\

Let us first consider a subamplitude's string with $n$ 1's
being decomposed into $m$ smaller strings, each with at least one 1.
This happens when, in the SD equation, an $(m+1)$-point vertex is encountered.
The number of inequivalent decompositions, denoted by $c_m(n)$, is
given by
\begin{equation}
c_m(n) = {1\over m!}\suml_{n_{1,2,\ldots,m}\ge1}
{n!\over n_1!\;n_2!\;\cdots\;n_m!}\;\;,  
\label{eq:compositions}
\end{equation}
where, of course, $n_1+n_2+\cdots+n_m=n$.
Note that the above equation assumes that all the subamplitudes containing
$n_1,n_2,\ldots,n_m$ external momenta exist. This is always the case when
a $\varphi^3$ interaction is present in the theory\footnote{because then there is always
the possibility of constructing a subamplitude containing $n_k$ external momenta
by combining a subamplitude containing $n_k -1$ momenta with an external momentum
in a three point vertex} but it is not true for a pure $\varphi^4$ theory for example. Then
one has to introduce a factor that cancels the terms coming from combinations of
non-allowed subamplitudes. We, nevertheless , proceed with our program to find a
generating function for effective theories that always contain a three point vertex.
We find
\begin{equation}
\suml_{n\ge0}{x^n\over n!}c_m(n) = {1\over m!}\left(e^x-1\right)^m\;\;.
\end{equation}
Now, out of all bit strings of size $K$, there are precisely
$K!/n!(K-n)!$ strings containing precisely $n$ 1's. The total number of
decompositions involving $(m+1)$-point vertices is therefore
\begin{equation}
f_m(K) =\suml_{n\ge0}\bino{K}{n}c_m(n)\;\;,
\end{equation}
so that
\begin{equation}
g_m(x) \equiv\suml_{K\ge0}{x^K\over K!}f_m(K) =
{1\over m!}e^x\left(e^x-1\right)^m\;\;.
\end{equation}
In the simple case of a pure $\f^3$ theory we therefore have
\begin{equation}
g_2(x) = {1\over2}\left(e^{3x}-2e^{2x}+e^x\right) =
\suml_{K\ge0}{x^K\over K!}\;{1\over2}\left(3^K-2^{K+1}+1\right)\;\;,
\end{equation}
so that the number of decompositions necessary to arrive at an $N$-point
amplitude is given by
$$
{1\over6}3^N-{1\over2}2^N+{1\over2}\;\;.
$$
For a theory with both $\f^3$ and $\f^4$ interactions such as gluonic QCD,
we find a total of
$$
{1\over24}4^N-{1\over4}2^N+{1\over3}
$$ decompositions.
In QCD at the tree level, an improvement is possible. We can decompose the
gluonic 4-vertex into two 3-vertices by employing an auxiliary field,
as explained for instance in \cite{petrosthesis}. This brings the
complexity down from $4^N$ to $3^N$, a worthwile improvement for large $N$.
It is not to be expected, however, that this will be possible in higher
orders. The effective action, therefore, will contain $(m+1)$-vertices for
all $m\ge2$, and the generating function is therefore
\begin{equation}
F(x) = \suml_{m\ge2}g_m(x) = \exp\left(e^x-1+x\right)-\exp(2x)\;\;.
\end{equation}
Below we give the number of decompositions,
\begin{equation}
D(N) = \suml_{m\ge2}g_m(N-1)\;\;,
\end{equation}
for not-too-large values of $N$.
\begin{center}
\begin{tabular}{|c|c|c|c|}\hline\hline
$N$ & $D(N)$ & $N$ & $D(N)$ \\ \hline
3 & 1 & 8 & 4,012 \\
4 & 7 & 9 & 20,891 \\
5 & 36 & 10 & 115,460 \\
6 & 171 & 11 & 677,550 \\
7 & 813 & 12 & 4,211,549 \\
\hline\hline
\end{tabular}\end{center}
For asymptotically large values of $N$, we have to study the analytic structure
of $F(x)$. Since this function is analytic for finite $x$, $D(N)$ must increase
with $N$ slower than $N!$. On the other hand, $D(N)$ increases faster than
$c^N$ for any finite $c$, which is reasonable since as $N$ grows larger and
larger values of $m$ come into play.
This is also evident from the fact that the standard Borel transform
of the series $F(x)$,
\begin{equation}
\int\limits_0^\infty dy\;e^{-y}\;F(xy) =
-{1\over1-2x} + \int\limits_0^\infty
dy\;\exp\left( -y + e^{xy} - 1 + xy \right)
\end{equation}
does not converge for any positive value of $x$.
\begin{center}
$$\plaat{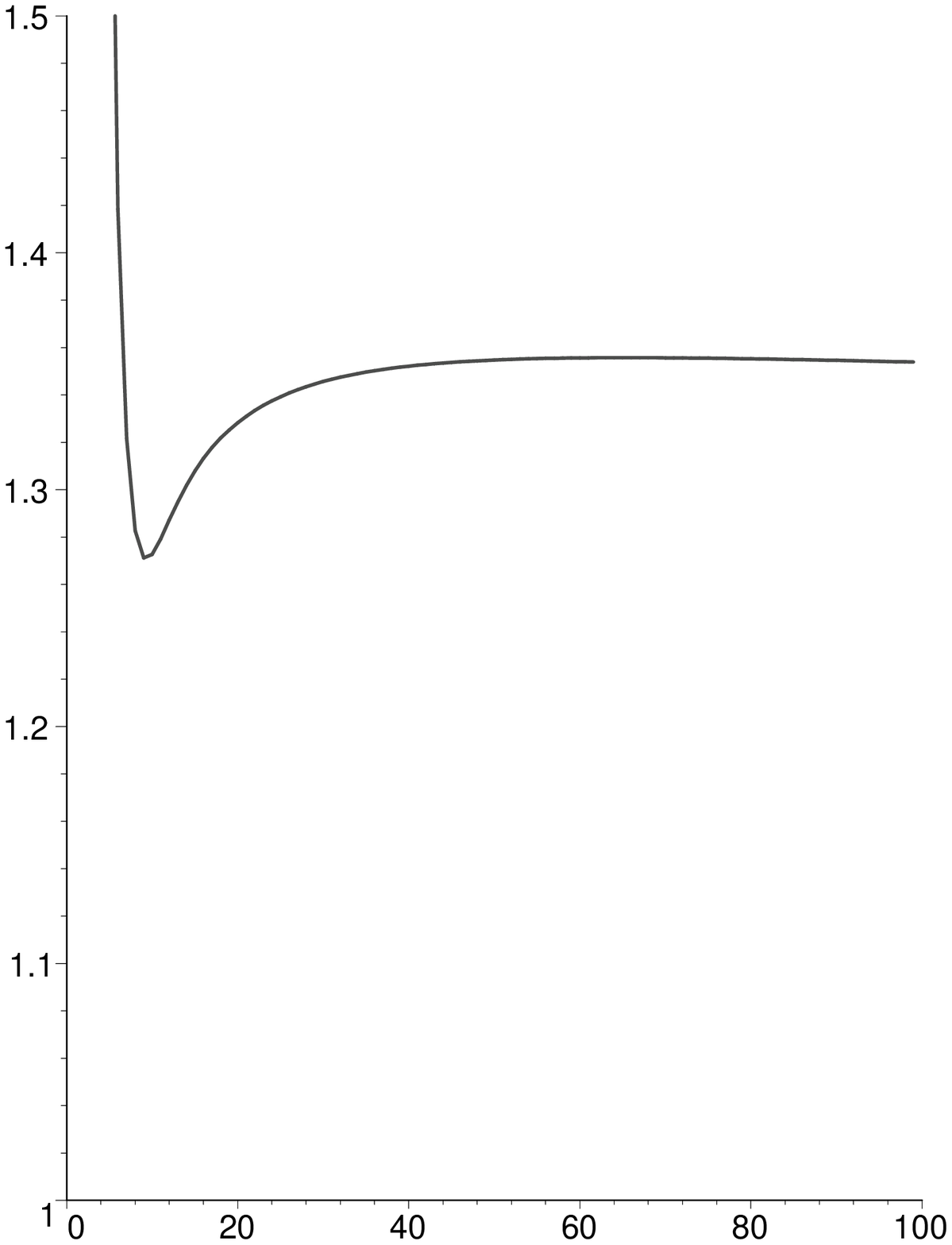}{8}{0}{8}$$
\end{center}
The above plot shows the behaviour of the ratio
$ (\log(N)/N)(D(N)/D(N-1))$
as a function of $N$ for $3\le N\le100$.
For high $N$, this ratio is approximately (but not quite) a constant.

\subsection{Complexity in one and two loops}

Consider a general theory  with $m$-point vertices.
Each subamplitude of level $n$ (containing $n$ specific external momenta), can be constructed
by combining two or more lower-level subamplitudes in a three- or more-point vertex.
When using an $(m+1)$-point vertex the subamplitude is built by $m$ lower level subamplitudes
and the number of different ways for this to happen is given by eq.~\ref{eq:compositions}.


Each term in the series represents the number of ways to construct the subamplitude
of level n using subamplitudes of level $n_1,\ldots,n_m$. The computational cost of each such subamplitude involves,
(assuming that there is an $m+1$ vertex in the theory)  contributions from the following posibilities:
All lower subamplitudes are free of loop corrections and the vertex is an ordinary one (this
gives the tree level subamplitude)\footnote{That is provided that the lower level subamplitudes exist!
This always happens when the theory involves $\varphi^3$ interactions. In the pure $\varphi^4$
theory, however, we have to modify the calculation to exclude combinations where one of the
$n_i$'s is equal to 2 since in such a theory there are no level 2 subamplitudes.}.
It can also be that one of the subamplitudes contains already a loop correction (that occured in
previous steps in the C.M. algorithm) and the vertex is an ordinary one (see fig ~\ref{eq:1lcm}). The subamplitudes
containing loop corrections can, however, be of level 2 or higher since the level one
subamplitudes are the external legs which we consider amputated.
There are, therefore, $m-\sum_i \delta_{1,n_i}$ different possibilities.
Finally there is the case that all subamplitudes are free of loop corrections but the vertex is actually a loop (see the last term in fig ~\ref{eq:1lcm}).
The number of different possibilities is now equal to the number of 1PI diagrams with one loop
and $m+1$ legs, which we denote with $J_{m,1}$.

\begin{equation}
\diagr{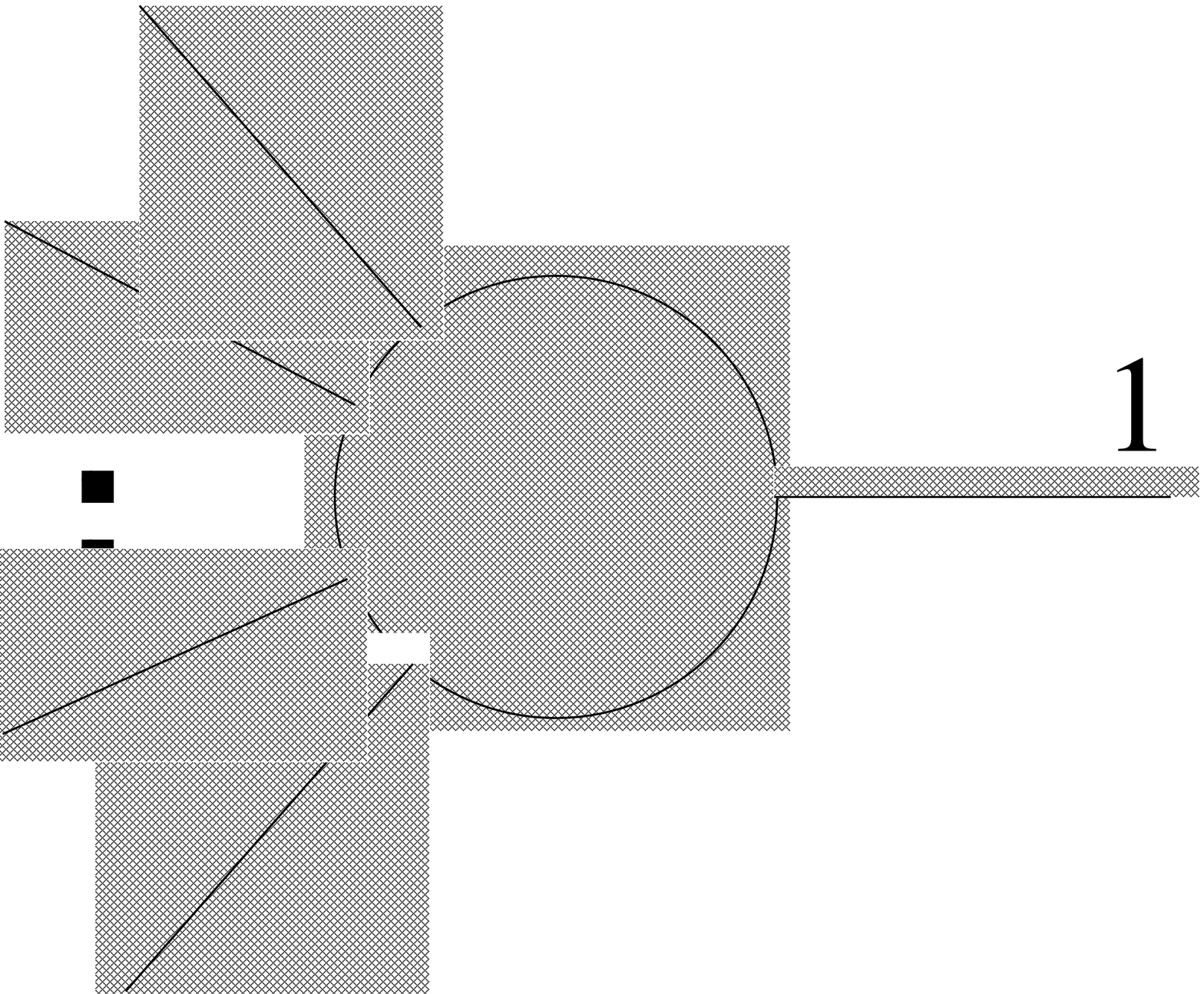}{1.3}{-.4}\;\;\;\;{=}\;\;\;
\;\diagr{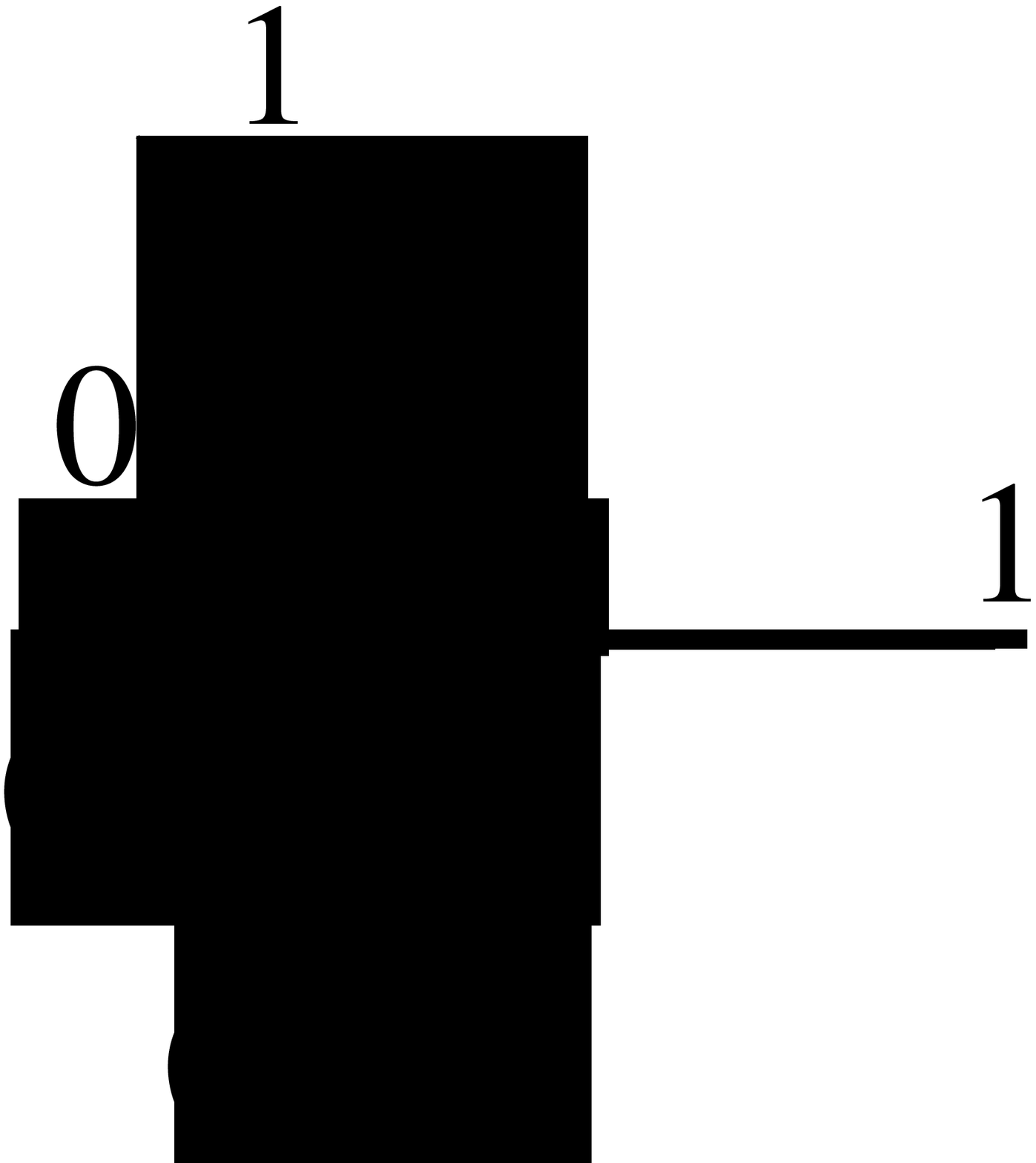}{1.3}{-.5}\;\;\;\;{+}\;\;\;
\;\diagr{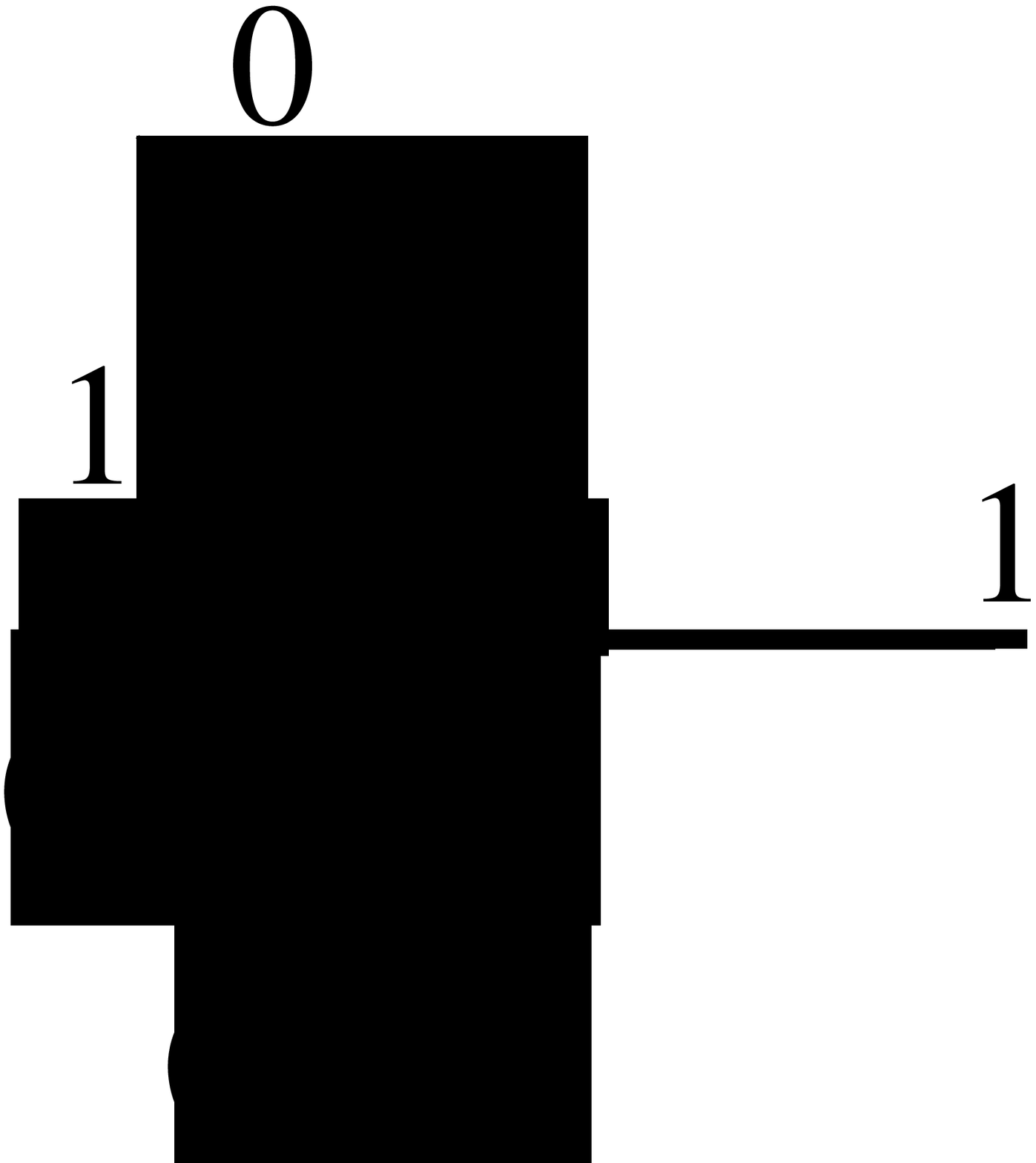}{1.3}{-.5}\;\;\;\;{+}\;\;\;\;{\ldots}\;\;\;\;{+}\;\;\;
\;\diagr{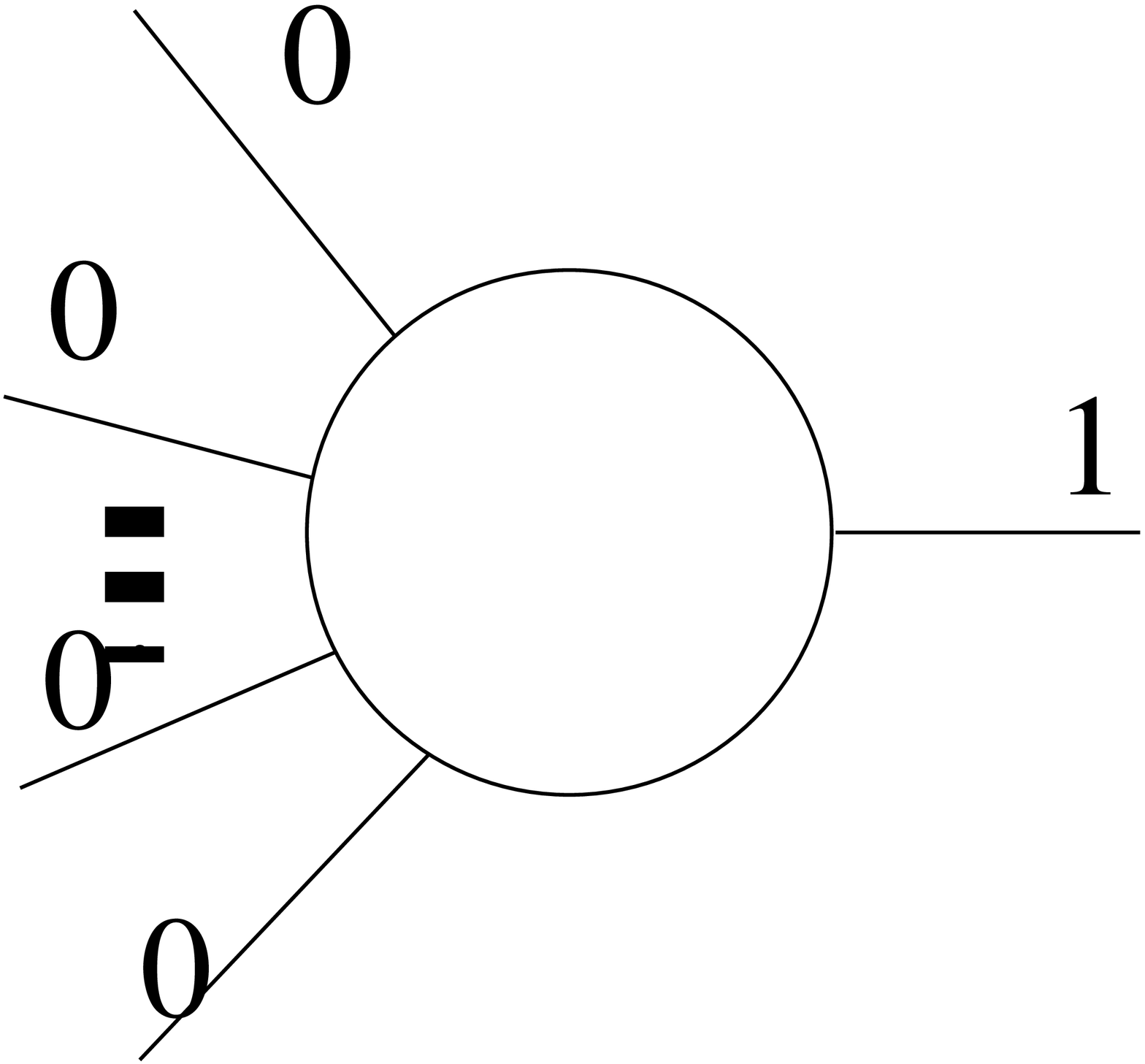}{1.3}{-.5}\;\;\;\;\;\;\;
\label{eq:1lcm}
\end{equation}

The cost of computing the specific subamplitude via an $m+1$ vertex is therefore

\begin{equation}
\frac{1}{m!} \sum_{n_1,\ldots,n_m}\frac{n!}{n_1!n_2!\ldots n_m!} \delta(\sum{n_i}-n) (V_m(1+m-\sum_i\delta_{1,n_i})+J_{m,1})
\eq

where we have included a factor $V_m=1$ if the $m+1$ vertex is in the theory and $V_m=0$ if not.
The cost of the subamplitude is then found by summing over $m$. There are $\frac{(N-1)!}{n!(N-1-n)!}$
different subamplitudes. The computational cost of the whole algorithm in units of effective vertices is then
\begin{equation}
\sum_{n=2}^{K}\bino{K}{n}\sum_{m \geq 2}^{\infty} \frac{1}{m!} \sum_{\stackrel{n_1,\ldots,n_m}{\sum n_i =n}}\frac{n!}{n_1!\ldots n_m!}  (V_m(1+m-\sum_i\delta_{1,n_i})+J_m)
\end{equation}
where $K=N-1$.  In the following table we present the results for the four
test theories.

\begin{center}
\begin{tabular}{|l|l|l|l|l|l|}
\hline \hline
  & $\varphi^3$ & $\varphi^4$ & $\varphi^3+\varphi^4$ & effective  \\
\hline
N=1 	& 0		& 0		& 0		& 0		\\
N=2  	& 0		& 0 		& 0 		& 0 			 \\
N=3  	& 2		& 0		& 5 		& 6 			 \\
N=4  	& 18		& 4 		& 46 		& 57 		\\
N=5  	& 114		& 0		& 340 		& 442 		\\
N=6  	&  720		& 105		& 2,715		& 3,713 		\\
N=7     & 5,368		& 0 		&26,346		&37,411		\\
N=8	&49,686		&3,395		&315,035	&459,056		\\
N=9	&553,766	&0		&4,474,868	&6,688,320		\\
N=10	&7,112,700	&149,140	&72,741,355	&112,139,709	\\
\hline
\hline
\end{tabular}
\end{center}

In order to include the two loop correction one has to add to the above
formula a term $(m-\sum \delta_{1,n_i})(m-\sum \delta_{n_i}-1)$ for the
possibility that two of the lower subamplitudes have a one loop correction and a term
equal to $m-\sum \delta_{1,n_i}$ for the possibility that one of the subamplitudes has a two loop correction.
There is also the possibility that one of the lower subamplitudes is of 1-loop order and
the vertex itself is a 1-loop 1PI graph. This costs an extra term of $J_{m,1}(m-\sum \delta_{1,n_i})$
Moreover one has to add the number of 1PI graphs with 2 loops and $m+1$ legs, $J_{m,2}$. Hence we
now have, writting $S_{n_i}=\sum_i \delta_{1,n_i}$

\begin{eqnarray*}
\sum_{n=2}^{K}\bino{K}{n}\sum_{m \geq 2}^{\infty} \frac{1}{m!} \sum_{\stackrel{n_1,\ldots,n_m}{\sum n_i =n}}\frac{n!}{n_1!\ldots n_m!}
(A+B+C)
\end{eqnarray*}

where $K=N-1$ and

\begin{eqnarray*}
A&=&V_m(1+m-S_{n_i}+(m-S_{n_i})(m-S_{n_i}-1)+m-S_{n_i})\\
B&=&J_{m,1}+J_{m,1}(m-S_{n_i})\\
C&=&J_{m,2}
\end{eqnarray*}

  The results for the four test cases are presented below:

\begin{center}
\begin{tabular}{|l|l|l|l|l|l|}
\hline \hline
  & $\varphi^3$ & $\varphi^4$ & $\varphi^3+\varphi^4$ & effective  \\
\hline
N=1 	& 0			& 0		& 0		& 0		\\
N=2  	& 0 			& 0 		& 0 		& 0 			 \\
N=3  	& 9 			& 0		& 45 		& 68			 \\
N=4  	& 102 			& 16 		& 566 		& 857 		\\
N=5  	& 957 			& 0		& 6,414 	& 9,837 		\\
N=6  	&  9,740		& 610		& 81,560	& 127,451 		\\
N=7     & 114,677		& 0 		&1,201,556	&1,920,824		\\
N=8	&1,546,986		&32,151		&20,211,345	&33,181,094		\\
N=9	&23,395,461		&0		&380,938,056	&644,468,452		\\
N=10	&390,310,512		&2,574,670	&7,929,937,496	&13,861,514,611	\\
\hline
\hline
\end{tabular}
\end{center}

One should be aware of the fact that the above results are obtained under the assumption that
the computational cost for every effective vertex that might include one or two loop 1PI
graphs is the same.

\section{Comparison of the complexity of the C.M. algorithm to the diagrammatic approach}

We present below the ratio of the computational complexity of the C.M. algorithm over
the number of diagrams one has to calculate in the customary diagrammatic approach,
for our four test theories\footnote{only amputated, tadpole/seagull-free diagrams are considered}.
 For each case the ratio for a calculation in tree, tree plus 1-loop, and
tree plus 1- and 2- loop level is presented.

\begin{center}
\begin{tabular}{|c|c|c|c|c|c|c|c|c|c|c|c|c|}
\hline \hline     
 \multicolumn{7}{|c|}{complexity of C.M. algorithm / number of diagrams}   \\
   \hline
  & \multicolumn{3}{c|}{$\varphi^3$} & \multicolumn{3}{c|}{$\varphi^4$}  \\
\hline
 & $L_0$& $L_0+L_1$ & $L_0+L_1+L_2$ &$L_0$ & $L_0+L_1$ & $L_0+L_1+L_2$\\
 \hline
N=3  	& 1.00 	& 1.000 & 1.500	& -	&- 	&-	 		\\
N=4  	& 2.00 	& 1.200 & 1.307	& 1.000 &1.000 	&1.231			\\
N=5  	& 1.666 & 0.864 & 0.955 & -	&- 	&-			\\
N=6  	& 0.857 & 0.516 & 0.679	& 2.00	&1.235 	&1.119			\\
N=7     & 0.318 & 0.309 & 0.498	& -	&- 	&-			\\
N=8	& 0.093 & 0.199 & 0.375	&1.575	&0.858 	&0.819			\\
N=9	& 0.022 & 0.136 & 0.286	&-	&- 	&-			\\
N=10	& 0.005 & 0.095 & 0.220	&0.636	&0.468 	&0.584		 	\\
\hline
\hline
\end{tabular}
\end{center}

\begin{center}
\begin{tabular}{|c|c|c|c|c|c|c|}
\hline \hline
   \multicolumn{7}{|c|}{complexity of C.M. algorithm / number of diagrams}   \\
   \hline
  & \multicolumn{3}{c|}{$\varphi^3+\varphi^4$} & \multicolumn{3}{c|}{effective}  \\
\hline
&$ L_0$& $L_0+L_1$ & $L_0+L_1+L_2$ &$L_0$ & $L_0+L_1$ & $L_0+L_1+L_2$\\
 \hline
N=3  	& 1.000 &1.000 &1.363 		& 1.000 &1.200 &1.619   \\
N=4  	& 1.500 &1.070 &1.132		& 1.500	&1.212 &1.325	\\
N=5  	& 1.00  &0.736 &0.828		& 0.961	&0.837 &0.956	\\
N=6  	& 0.409	&0.451 &0.605		& 0.381	&0.519 &0.691	\\
N=7     & 0.121	&0.283 &0.454		&0.109	&0.327 &0.515	\\
N=8	&0.028	&0.190 &0.347		&0.025  &0.217 &0.392	\\
N=9	&0.005	&0.132 &0.268		&0.005  &0.150 &0.302	\\
N=10	&0.001  &0.094 &0.208		&0.001  &0.107 &0.234	\\
\hline
\hline  

\end{tabular}
\end{center}

One should note that the C.M. algorithm will actually perform better than depicted by the above numbers,
when compared with the straightforward diagrammatic approach, since we consider the cost of a step in the C.M. algorithm
(i.e. the calculation of a subamplitude which corresponds to the calculation of an effective vertex)  equal to the cost of the computation
of a whole diagram. That is the reason for the apparently poor performance of the C.M. algorithm in the case of tree level $\varphi^4$ theory.

We, therefore, conclude that the Caravaglios-Moretti algorithm is more effective than the straightforward diagrammatic approach,
in the tree as well as the one and two loop level, by a factor that increases rapidly with the number of external legs, even though this increase
is less rapid in the one- and two- loop level than in tree level.

\end{document}